\documentclass[mathleft]{an}
\usepackage{graphicx}
\usepackage{times}
\def\degr{\hbox{$^\circ$}}
\def\arcmin{\hbox{$^\prime$}}
\def\arcsec{\hbox{$^{\prime\prime}$}}

\begin{document}

\Pagespan{548}{}
\Yearpublication{2013}%
\Yearsubmission{2013}%
\Month{07}%
\Volume{334}%
\Issue{6}%
\DOI{10.1002/asna.201311894}%

\title{The LOFAR View of Cosmic Magnetism}

\author{R. Beck\inst{1}\thanks{On behalf of the LOFAR collaboration. Corresponding author:
  \email{rbeck@mpifr-bonn.mpg.de}\newline}
\and J. Anderson\inst{1}
\and G. Heald\inst{2}
\and A. Horneffer\inst{1} \and M. Iacobelli\inst{3} \and J.
K\"ohler\inst{1} \and D. Mulcahy\inst{1} \and R. Pizzo\inst{2}
\and A. Scaife\inst{4}
\and O. Wucknitz\inst{1,5}
\and the team of the LOFAR Magnetism Key Science Project}
\titlerunning{The LOFAR view of Cosmic Magnetism}
\authorrunning{R. Beck et al.}
\institute{MPI f\"ur Radioastronomie, Auf dem H\"ugel 69, 53121 Bonn, Germany
\and ASTRON, Postbus 2, 7990
AA Dwingeloo, The Netherlands
\and Leiden Observatory, Leiden University, PO Box
9513, 2300 RA Leiden, The Netherlands
\and School of Physics \& Astronomy, University of Southampton, Southampton, Hampshire, SO17 1BJ, United Kingdom
\and Argelander-Institut f\"ur Astronomie, Auf dem H\"ugel 71, 53121 Bonn, Germany}

\received{2013 Feb 8} \accepted{2013 Apr 15}


\abstract{%
The origin of magnetic fields in the Universe is an open problem in
astrophysics and fundamental physics. Polarization observations with
the forthcoming large radio telescopes will open a new era in the
observation of magnetic fields and should help to understand their
origin. At low frequencies, LOFAR (10--240~MHz) will allow us to map
the structure of weak magnetic fields in the outer regions and halos
of galaxies, in galaxy clusters and in the Milky Way via their
synchrotron emission. Even weaker magnetic fields can be measured at
low frequencies with help of Faraday rotation measures. A detailed
view of the magnetic fields in the local Milky Way will be derived
by Faraday rotation measures from pulsars. First promising images
with LOFAR have been obtained for the Crab pulsar-wind nebula, the
spiral galaxy M~51, the radio galaxy M~87 and the galaxy clusters
A~2255 and A~2256. With help of the polarimetric technique of
``Rotation Measure Synthesis'', diffuse polarized emission has been
detected from a magnetic bubble in the local Milky Way. Polarized
emission and rotation measures were measured for more than 20
pulsars so far.}

\maketitle

\section{Origin of magnetic fields}

Cosmic magnetic fields belong to the {\em Dark Side of the
Universe}. The physical processes of their generation and evolution
are widely unknown. Observational data are hard to obtain because
cosmic magnetic fields need ``illumination'', e.g. by cosmic-ray
electrons. Magnetic fields in the interstellar medium are strong
enough to be dynamically important, but the resolution of
present-day radio telescopes is insufficient to study the
interaction between gas and magnetic fields in detail. Magnetic
fields form halos around galaxies and galaxy clusters which may
extend much further than known today. Even intergalactic space may
be magnetized, but an observational proof is extremely difficult.

The origin of the first magnetic fields in the early Universe is
particularly mysterious (\cite{widrow02}). Any large-scale
primordial field is hard to maintain in a young galaxy because
differential rotation winds up the field lines during galaxy
evolution and the field is rapidly destroyed by reconnection.
Small-scale ``seed'' fields could originate from the time of
cosmological structure formation by the Weibel instability (a
small-scale plasma instability) (\cite{lazar09}) or from injection
by the first stars or jets generated by the first black holes (\cite
{rees05}). The most promising mechanism to sustain magnetic fields
in the interstellar medium of galaxies is the dynamo
(\cite{beck96,gent12}). The magnetic fields in the gaseous halos of
galaxy clusters could be seeded by outflows from starburst galaxies
(\cite{donnert08}) or from AGNs, and amplified by turbulent wakes,
cluster mergers or a turbulent dynamo (\cite{ryu08}).

The Low Frequency Array (LOFAR), the planned Square Kilometre Array
(SKA) (Sect.~\ref{ska}) and its precursor telescopes under
construction, the Australia SKA Pathfinder (ASKAP) and the Karoo
Array Telescopes (MeerKAT) in South Africa, are large
next-generation radio telescopes with high sensitivity and high
resolution. They will open a new era in studying cosmic magnetism.

\section{The advantages of low-frequency radio astronomy}
\label{rm}

\begin{figure}
\centering
\includegraphics[width=0.48\textwidth]{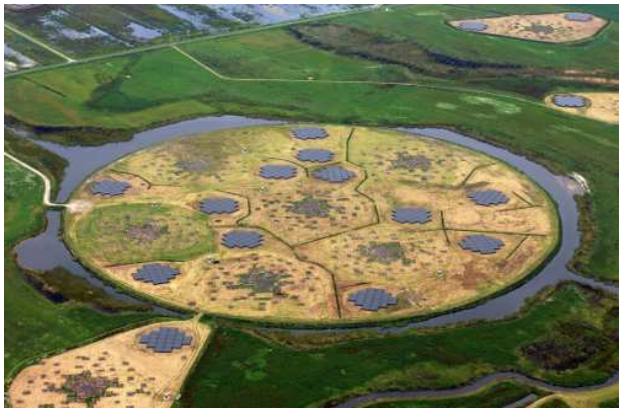}
\caption{The inner LOFAR core near Exloo (Netherlands) (copyright:
ASTRON).} \label{fig:lofar}
\end{figure}

The low-frequency regime bears surprises, as much of this regime has
been `terra incognita' until recently. Low-frequency radio waves
from the Milky Way and galaxies are synchrotron emission from
cosmic-ray electrons, with only a small admixture of thermal
emission. Diffuse radio emission from halos of galaxy clusters is
purely of synchrotron origin. Its intensity and polarization can be
used to measure the strength and morphology of the interstellar and
intergalactic magnetic fields. This implies several advantages in
using radio telescopes at low frequencies. The power-law nature of
synchrotron radiation with a negative spectral index means that the
intensity increases towards decreasing frequencies. Furthermore,
measurement of diffuse, faint emission is easier than at higher
frequencies, where the lifetime of electrons is limited by various
energy loss processes, so that the extent of synchrotron sources is
smaller and the synchrotron spectrum is steeper.

Synchrotron emission at low frequencies is still observable from
aging, low-energy electrons that have propagated far away from their
places of origin. Hence, LOFAR is a suitable instrument to search
for weak magnetic fields in outer galaxy disks, galaxy halos and
halos of galaxy clusters (Sect.~\ref{mksp}). ``Relics'' in galaxy
clusters, probably signatures of giant shock fronts, have steep
radio spectra and become prominent in low frequencies
(\cite{brunetti08,weeren12b}).

The intrinsic degree of linear polarization of synchrotron emission
is about 75\%. The observed degree of polarization is smaller due to
the contribution of unpolarized thermal emission, which is weak at
low frequencies, by Faraday depolarization along the line of sight
and across the telescope beam (\cite{sokoloff98}) and by geometrical
depolarization due to field bending within the beam. Faraday
depolarization also occurs when polarized emission is averaged over
a large bandwidth. As Faraday depolarization increases with
decreasing frequency, polarized intensity is reduced and may reveal
a positive spectral index at low frequencies (\cite{arshakian11}).
Polarized emission cannot be detected from the star-forming disks of
galaxies, but from regions in the outer disk or halo where the
density of thermal electrons is low, magnetic fields are weak and
hence Faraday depolarization is small.

Another important aspect of low-frequency observations is their
sensitivity to small Faraday depths. The Faraday rotation angle
$\Delta\chi$ is proportional to the ``Faraday depth'' (FD), defined
as the line-of-sight integral over the product of the plasma density
and the strength of the field component along the line of sight.
Because the Faraday rotation angle increases with the square of
wavelength $\lambda^2$, low frequencies are ideal to search for
small Faraday depths from weak interstellar and intergalactic
fields.

\begin{figure}
\centering
\includegraphics[width=0.4\textwidth]{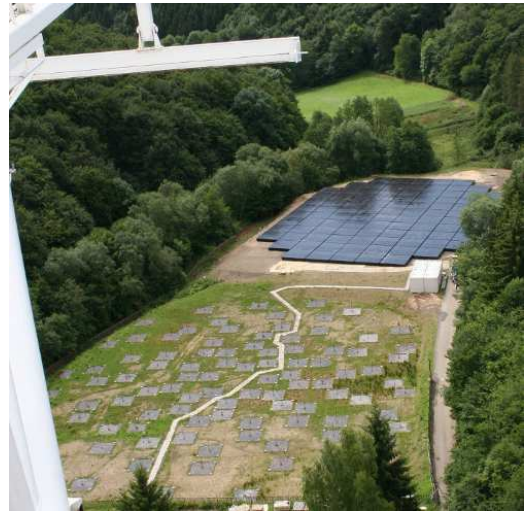}
\caption{International LOFAR station seen from the lower platform of
the Effelsberg 100-m telescope. Front: sparse dipole array for
10--80\,MHz; back: dense array of dipole ``tiles'' for frequencies
of 110--240\,MHz (copyright: James Anderson, MPIfR).}
\label{fig:effelsberg}
\end{figure}

\begin{figure}
\centering
\includegraphics[width=0.45\textwidth]{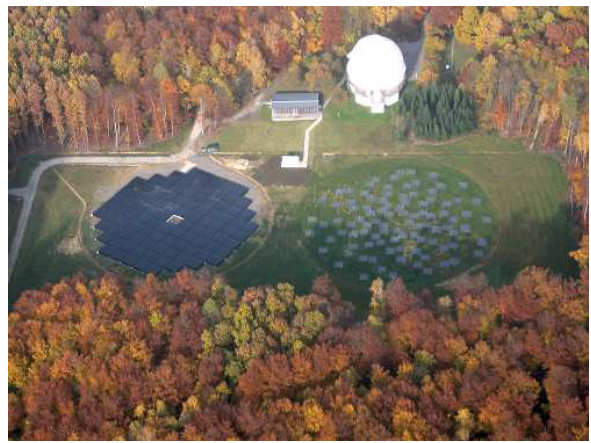}
\caption{International LOFAR station near Tautenburg/Germany. Front
left: dense array of dipole ``tiles'' for frequencies of
110--240\,MHz; front right: sparse dipole array for 10--80\,MHz;
back: optical Schmidt telescope (copyright: Michael Pluto,
Th\"uringer Landessternwarte).} \label{fig:tautenburg}
\end{figure}

Modern radio telescopes are equipped with digital correlators that
allow us to record a large number of spectral channels. While radio
spectroscopy in total intensity is well developed, the possibilities
of spectro-polarimetry in radio continuum are explored only since a
few years. The new method of ``Rotation Measure Synthesis'' applied
to multi-frequency polarization data generates the ``Faraday
dispersion function'' or, in short, the ``Faraday spectrum''
(\cite{brentjens05,bell11,frick11}). Multiple emitting and rotating
regions located along the line of sight generate several components
in the Faraday spectrum. As in classical spectroscopy, the
interpretation of this spectrum is not straightforward. In
particular, there is no simple relation between Faraday depth and
geometrical depth. Only in case of a single component in the
spectrum, its Faraday depth is identical to the classical ``Faraday
rotation measure''. A large span in $\lambda^2$ covered by
observations implies a high resolution in Faraday spectra
(\cite{brentjens05,heald09,beck+12}), with which any foreground,
intermittent or source-intrinsic features in Faraday spectra can be
identified more easily.

A next major step to improve the quality of polarization images is
to combine image synthesis with RM Synthesis to ``Faraday
Synthesis'', which will soon be applied to 3-D data cubes
(\cite{bell12}).

In summary, the advantages of low-frequency observations are:

\begin{itemize}
\item New sources with ultra-deep spectra may be discovered.
\item The emission is almost purely synchrotron (no thermal
admixture).
\item The intensity is high because synchrotron spectra are generally
steep.
\item The lifetime of cosmic-ray electrons is higher and hence the
extent of synchrotron sources is larger.
\item Faraday rotation angles are larger, so that smaller rotation
measures (or Faraday depths, to be precise) can be measured.
\item A large coverage in $\lambda^2$ and hence high resolution in Faraday
spectra can be achieved with one single telescope.
\item Antennas and receiving systems are relatively cheap.
\end{itemize}

\section{The Low Frequency Array} \label{lofar}

\begin{figure}
\centering
\includegraphics[width=0.45\textwidth]{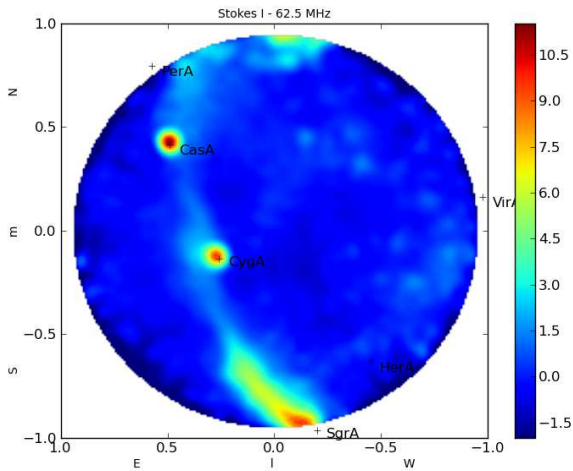}
\caption{All-sky image at 62.5\,MHz, obtained on 22 Aug 2011 from
1.3\,sec observation time with the Effelsberg LOFAR station. The
angular resolution is about 6\degr. The brightest radio sources
Cas~A, Cyg~A and Sgr~A are marked (from Jana K\"ohler and James
Anderson, MPIfR).} \label{fig:sky}
\end{figure}

The meter-wave radio telescope LOFAR (Low Frequency Array, {\em
www.lofar.org}) was designed by ASTRON, the Netherlands Institute
for Radio Astronomy. With its huge collecting area it is the largest
online connected radio telescope ever built (\cite{brueggen06}).
LOFAR is a software telescope with no moving parts, steered solely
by electronic phase delays. It has a huge field of view and can
observe towards many directions simultaneously. LOFAR is an
interferometric array using about 43\,000 small antennas of two
different designs, one for the wavelength range 10--80\,MHz (``low
band'') and one for 110--240\,MHz (``high band'').

The antennas are aggregated of at least 49 stations with baselines
up to more than 1000\,km across Europe. 34 of these stations are
distributed across the Netherlands (plus six following in 2013),
five stations in Germany (plus another one following in 2013), one
each in Great Britain, France and Sweden, which are jointly operated
as the International LOFAR Telescope (ILT). Another three stations
are planned in Poland and further stations may also be built in
other European countries. The core stations are located about 3\,km
north of the village of Exloo in the Netherlands
(Fig.~\ref{fig:lofar}). The angular resolution of a single
international station is between 2\degr\ at 240\,MHz and about
23\degr\ at 15\,MHz. The total effective collecting area is up to
approximately 300\,000\,m$^2$, depending on frequency and antenna
configuration. The angular resolution is between 2\arcsec\ (at
240\,MHz) and 15\arcsec\ (at 30\,MHz) for the Dutch stations (up to
about 100\,km baselines), but can be improved by about 10x when the
international baselines are included. Digital beam forming allows
observation towards several directions simultaneously. The data
processing is performed by a supercomputer situated at the
University of Groningen/Netherlands.

The first German station was built in 2007 next to the 100-m
Effelsberg radio telescope and completed in July 2009
(Fig.~\ref{fig:effelsberg}), the second near Tautenburg
(Th\"uringen) (Fig.~\ref{fig:tautenburg}) was completed in November
2009, the third German stations near Garching (Unterweilenbach) in
2010, the fourth and fifth stations in Bornim near Potsdam and in
J\"ulich in 2011. Another station in Hamburg will be built in 2013.
The 12 participating German institutes are organized in GLOW (German
Long Wavelength Consortium). Their main scientific interests are the
Epoch of Reionization, when the first cosmic gas structures formed,
magnetic fields and cosmic rays in our Milky Way, in galaxies and in
jets, pulsars and solar radio emission (see {\em www.lofar.de}).

LOFAR was officially opened by the Dutch Queen on 12 June 2010.
Regular science observations with 33 Dutch stations, five German
stations and the stations in the UK and France started in December
2012.

\begin{figure}
\centering
\includegraphics[width=0.4\textwidth]{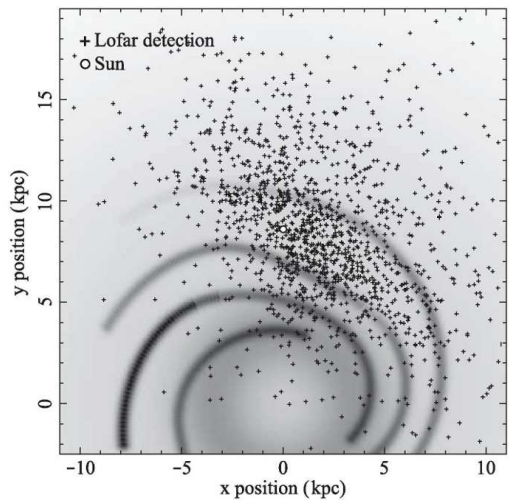}
\caption{Simulation of the 1000+ pulsars that LOFAR is expected to
find in a 60-day all-sky survey, shown in a Galactic plane
projection (from \cite{leeuwen10}). } \label{fig:pulsars1}
\end{figure}

The sensitivities and spatial resolutions attainable with LOFAR will
allow several fundamentally new studies:

\begin{itemize}

\item search for the signature of the reionization of
neutral hydrogen in the distant Universe ($6 < z < 10$), making use
of the shift of the 21\,cm line into the LOFAR high band window;
\item detect the most distant massive galaxies and
study the processes of formation of the earliest structures in the
Universe: galaxies, galaxy clusters and active galactic nuclei;
\item map the three-dimensional distribution of magnetic
fields in our own and nearby galaxies, in galaxy clusters and in the
intergalactic medium (Sect.~\ref{science});
\item discover about 1000 new pulsars within a few
kiloparsecs from the Sun (Fig.~\ref{fig:pulsars1});
\item detect flashes of low-frequency radiation from the outer solar
planets and from pulsars, search for bursts from Jupiter-like
extrasolar planets and for short-lived transient events produced by
e.g. stellar mergers or black hole accretion;
\item detect ultra-high energy cosmic rays entering
the Earth's atmosphere;
\item detect coronal mass ejections from the sun and
provide large-scale maps of the solar wind;
\item explore the low-frequency spectral window, to make
unexpected ``serendipitous'' discoveries.

\end{itemize}

\section{Prospects of magnetic field observations with LOFAR}
\label{science}

\subsection{Jupiter's magnetosphere}

\begin{figure}
\centering
\includegraphics[width=0.4\textwidth]{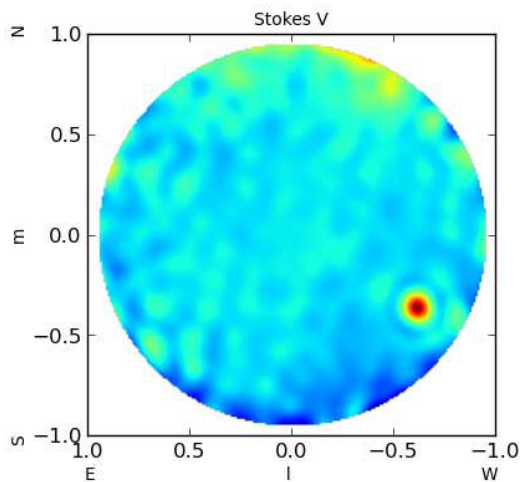}
\caption{All-sky Stokes V image at 32\,MHz with Jupiter on bottom
right, obtained on 3 Aug 2012 from 1.3\,sec observation time with
the Dutch LOFAR station CS002. The degree of circular polarization
is 57\% at this frequency (from Jana K\"ohler and James Anderson,
MPIfR).} \label{fig:jupiter}
\end{figure}

The outer gas planets of the solar system are strong radio emitters,
producing by various processes diverse radio components with complex
dynamic spectra (\cite{zarka04a}). The most intense radio emissions
of Jupiter below about 40\,MHz are related to the aurorae and the
moon Io and are generated by electrons accelerated to keV energies
in the magnetosphere, via a well studied nonthermal coherent
process: the Cyclotron Maser Instability (CMI). This emission is
strongly elliptically polarized. Jupiter is the brightest
low-frequency source in the sky in total intensity and by far the
dominating source in Stokes V (Fig.~\ref{fig:jupiter}). Intense
synchrotron emission (incoherent and thus less intense than auroral
emission) is also produced from MeV electrons trapped in Jupiter's
radiation belts. Atmospheric lightning is accompanied by the
emission of broadband radio bursts (\cite{zarka04c}).

\begin{figure}
\centering
\includegraphics[width=0.45\textwidth]{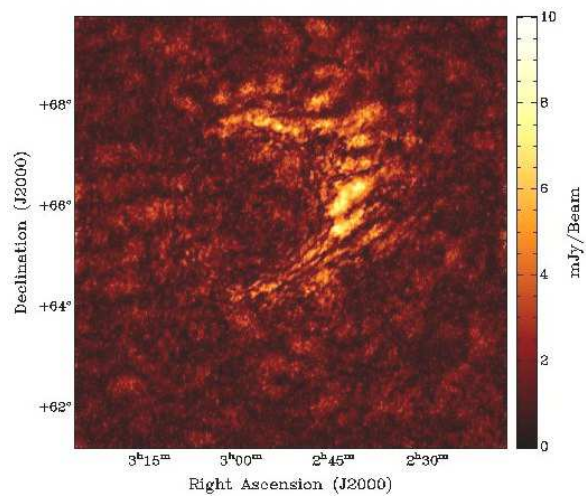}
\caption{A region of 15\degr\ x 15\degr\ in the ``Fan'' region of
the Milky Way's plane centered at $l=137\degr, b=+7\degr$ was
observed with LOFAR at 110--174\,MHz with 139\arcsec\ x 126\arcsec\
resolution. The polarized intensity of a slice through the Faraday
data cube at Faraday depth -2\,rad/m$^2$ is shown (from Marco
Iacobelli, University of Leiden, and the MKSP commissioning team).}
\label{fig:fan1}
\end{figure}

\begin{figure}
\centering
\includegraphics[width=0.45\textwidth]{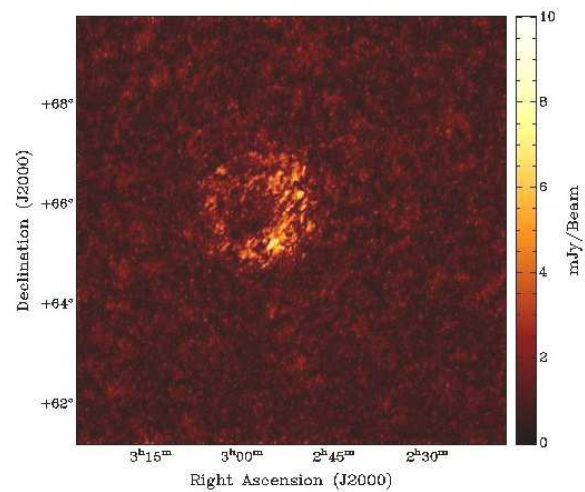}
\caption{Same as Fig.~\ref{fig:fan1}, but at Faraday depth
-5\,rad/m$^2$ (from Marco Iacobelli, University of Leiden, and the
MKSP commissioning team).} \label{fig:fan2}
\end{figure}

Detailed studies of Jupiter planned with LOFAR (\cite{zarka04b}) are
crucial for the extrapolation to radio emissions from exoplanets.
Dynamic spectra have already been studied with single stations and
with the longest LOFAR baselines and will be used soon to
investigate the position and motion of the emitting electrons
(Wucknitz \& Zarka, in prep.).

\subsection{Polarization of the Galactic foreground}

LOFAR's broad coverage of low frequencies makes it uniquely suited
to studying weak magnetic fields and low-density regions in the halo
of the Milky Way, where such studies can address both the disk-halo
interaction and the energetics of the interstellar medium. Emission
and Faraday rotation from magnetic fields in the Galaxy affect the
measurement of distant, extended objects. At LOFAR frequencies this
foreground influence must be properly separated for a correct
interpretation of data from extragalactic sources. RM Synthesis
(Sect.~\ref{rm}) of diffuse Galactic synchrotron emission is an
excellent way to disentangle various components in the Faraday
spectrum, giving statistical information on the clumped magnetized
ISM and on the relation of thermal electron density to magnetic
field strength. Data from regions with a minimum of Galactic
foreground contamination will be used to make a statistical
comparison of RMs and redshifts in order to investigate the presence
and properties of magnetic fields in the Galactic halo.

Faraday depolarization is strong at low frequencies. Polarized
emission from the Milky Way can be detected only from nearby
magnetized objects. These rotate the polarization angle of
background emission and hence act as a ``Faraday screen''. The
Canadian/German Galactic Plane Survey (\cite{landecker10}) revealed
a wealth of structures in polarized emission which are invisible in
total intensity.

One of the first detections of polarized emission with LOFAR was
achieved from a huge gas bubble in the ``Fan'' region of the Milky
Way (\cite{iacobelli13}). Figs.~\ref{fig:fan1} and \ref{fig:fan2}
show slices through the Faraday data cube at two different Faraday
depths, corresponding to two different geometrical depths. The ring
is part of a magnetized bubble structure, possibly a relic
(\cite{iacobelli+13}).

\subsection{Large-scale magnetic fields of the Milky Way}
\label{milkyway}

\begin{figure}
\centering
\includegraphics[width=0.48\textwidth]{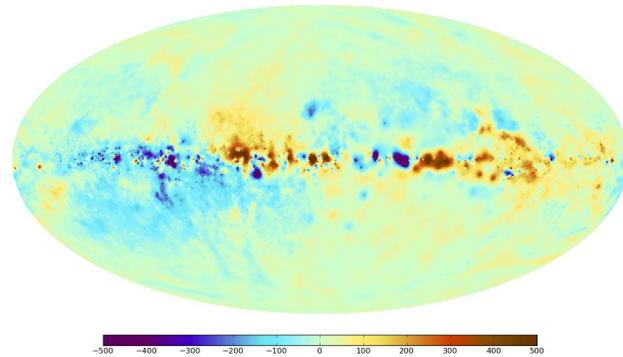}
\caption{Model of the RM sky looking through the Milky Way (in
Galactic coordinates), based on 41\,330 RMs from polarized
background sources (from \cite{oppermann12}).} \label{fig:rm}
\end{figure}

Faraday rotation measures (RMs) of polarized background sources
reveal an image of the regular magnetic fields averaged along the
lines of sight. RMs are positive around $l\approx250\degr$ and
negative around $l\approx80\degr$ (Fig.~\ref{fig:rm}), so that the
magnetic field of the local spiral arm is directed clockwise and is
symmetric with respect to the Galactic plane. The different RM signs
above and below the Galactic plane around $l\approx30\degr$ may
indicate an antisymmetric field component towards the inner galaxy
or in the Galactic halo, but this could also be the effect of a
nearby magnetized cloud.

\begin{figure}
\centering
\includegraphics[width=0.35\textwidth]{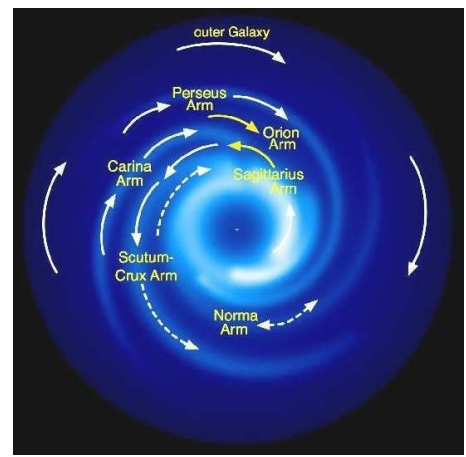}
\caption{Sketch of the magnetic field in the Milky Way, as derived
from Faraday rotation measures of pulsars and extragalactic sources.
Generally accepted results are indicated by yellow vectors, while
white vectors refer to results which need confirmation (from Jo-Anne
Brown, Calgary).} \label{fig:galaxy}
\end{figure}

Pulsars play a central role in detecting interstellar magnetic
fields in the Galaxy. RM surveys of polarized pulsars were used to
model the large-scale regular field of the Milky Way
(\cite{eck11,noutsos12}). One large-scale field reversal is required
at about 1--2\,kpc from the Sun towards the Milky Way's center
(Fig.~\ref{fig:galaxy}); more reversals possibly exist but cannot be
confirmed with the present data. Nothing similar has yet been
detected in other spiral galaxies, although high-resolution RM maps
of Faraday rotation are available for many spiral galaxies.
Large-scale field reversals may be relics of seed fields from the
early phase of a galaxy which were stretched by differential
rotation (\cite{moss12}).

\begin{figure}
\centering
\includegraphics[width=0.4\textwidth]{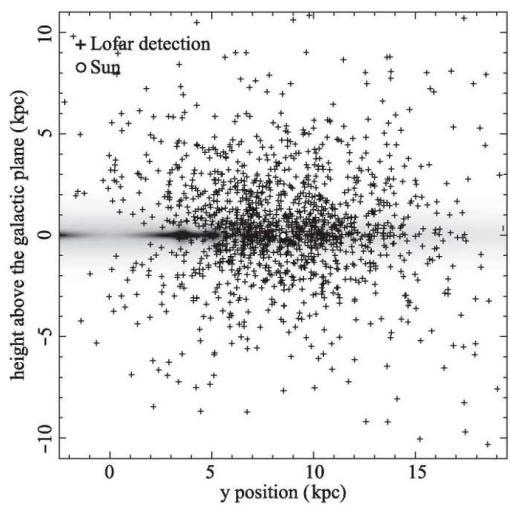}
\caption{Same as Fig.~\ref{fig:pulsars1}, but as an edge-on view of
the Galactic plane.} \label{fig:pulsars2}
\end{figure}

LOFAR pulsar searches will benefit from both high sensitivity and an
increasing pulsar brightness at low frequencies. This is expected to
result in the discovery of a new population of dim, nearby and
high-latitude pulsars too weak to be found at higher frequencies.
LOFAR will detect almost all pulsars within 2\,kpc of the Sun; at
least 1\,000 pulsar discoveries are expected from LOFAR, especially
at high latitudes (Fig.~\ref{fig:pulsars2}). Most of these are
expected to emit strong, linearly polarized signals and will allows
us to measure their RMs. This should approximately double the
current RM sample ($\approx 700$ RMs). When combined with the
catalogue of extragalactic-source RMs (Fig.~\ref{fig:rm}), this will
provide the strength and direction of the regular magnetic field in
previously unexplored directions and locations in the Galaxy.

Very little is known about the magnetic field properties of the
Milky Way beyond a few hundred parsecs from the Galactic plane. RMs
of high-latitude pulsars and extragalactic sources are crucial for
determining fundamental properties such as the scale height and
geometry of the magnetic field in the thick disk and halo, as well
as providing the exciting prospect of discovering magnetic fields in
globular clusters.

In addition to providing an indirect probe of Galactic magnetic
structure, polarization surveys of pulsars at low frequencies will
allow to study the effects of scattering in the interstellar medium,
which are prominent at LOFAR frequencies but have until now mostly
been studied at higher frequencies (\cite{noutsos09}); pulsar
polarization spectra, which are expected to turn-over in the LOFAR
high band, and constraints on the geometry of pulsar magnetospheric
emission.

\begin{figure}
\centering
\includegraphics[width=0.45\textwidth]{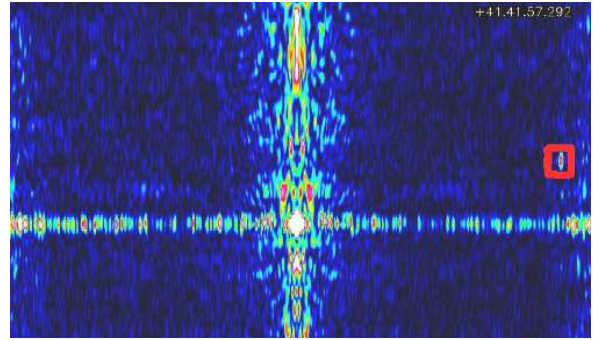}
\caption{Slice through a RM Synthesis cube at the declination of the
pulsar PSR\,J0218+42, made from 5\,min worth of LOFAR data. The plot
has Faraday depth (FD) on the x-axis ($\pm70$\,rad\,m$^{-2}$) and
right ascension on the y-axis. The pulsar at a FD of
61\,rad\,m$^{-2}$ is highlighted by a red box. At around
0\,rad\,m$^{-2}$ and at the position of a strong source at
RA=2h\,34m\,30s there is significant instrumental polarization
(vertical and horizontal stripes). With improved calibration the
instrumental polarization will be reduced, but already at this stage
the pulsar can be easily picked out (from A. Horneffer, MPIfR Bonn,
and M.\,R. Bell, MPA Garching).} \label{fig:J0218}
\end{figure}

Polarized emission and rotation measures have been measured with the
LOFAR pulsar observation mode for more than 20 pulsars so far (Sobey
\& Noutsos, in prep). Pulsars are also visible in Faraday spectra of
imaging observations as strongly polarized sources
(Fig.~\ref{fig:J0218}).

\subsection{Magnetic fields in the Crab Nebula}

\begin{figure}
\centering
\includegraphics[width=0.35\textwidth]{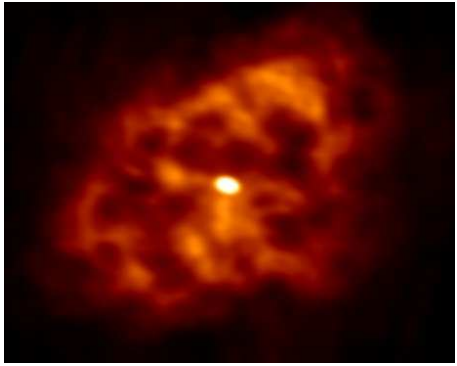}
\caption{Total synchrotron intensity of the Crab nebula, observed
with the Dutch LOFAR stations in the high band (115--150\,MHz). The
resolution is 9\arcsec\ x 14\arcsec, the field size is 7.5\arcmin\ x
6\arcmin\ (from Olaf Wucknitz, MPIfR).} \label{fig:crab}
\end{figure}

The Crab pulsar-wind nebula was one of the early targets of LOFAR
observations in February 2011 (Fig.~\ref{fig:crab}). The structure
is very similar to that at higher frequencies (\cite{bietenholz90}),
so that the radio spectral index is constant across the nebula. This
suggests that the dominant accelerator of the relativistic electrons
is the central pulsar. The magnetic field strengths in the filaments
are not known, probably a few 100~$\mu$G (\cite{bietenholz91}).

Observing with the long baselines of the international LOFAR
stations, the radio signal is dominated by the pulsar, which served
as a point-like calibrator source. This calibration was then used to
image the shorter baselines from the Dutch stations to reveal the
structure of the nebula. The long baseline data themselves allowed
to study the polarization of the pulsar, which turned out to be much
lower than in earlier observations with the Westerbork telescope at
similar frequencies (Wucknitz et al., in prep).

\subsection{Magnetic fields in spiral galaxies}
\label{galaxies}

It is now generally accepted that galactic magnetic fields result
from the amplification of a seed magnetic field by a hydromagnetic
dynamo, rather than having a merely primordial origin. Turbulent
``seed'' fields in young galaxies can originate from the Weibel
instability in shocks during the cosmological structure formation
(\cite{lazar09}), injected by the first stars or jets generated by
the first black holes (\cite{rees05}). In young galaxies a
small-scale dynamo possibly amplified the seed fields from the
protogalactic phase to the energy density level of turbulence within
less than $10^8$\,yr (\cite{schleicher10,abeck12}), followed by a
large-scale dynamo which generates a spiral field pattern
(\cite{beck96,hanasz09,arshakian09}). Field seeding by supernova
explosions and amplification of small-scale fields operates over the
whole lifetime of a galaxy and may affect the regularity of the
field pattern (\cite{moss12}).

\begin{figure}
\centering
\includegraphics[width=0.36\textwidth]{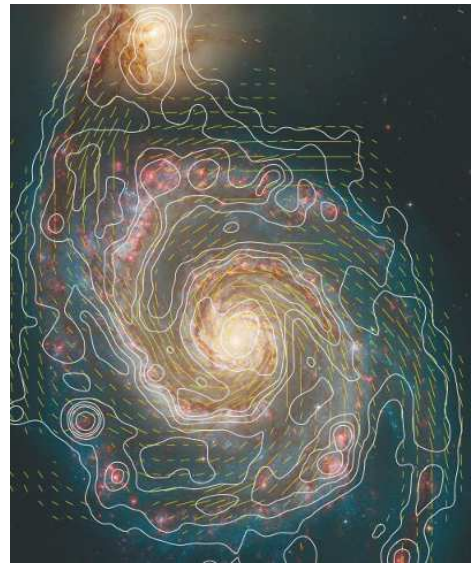}
\caption{Total radio intensity (contours) and B--vectors of M~51,
combined from observations at 4.8\,GHz with the VLA and Effelsberg
telescopes and smoothed to 15\arcsec\ resolution, overlaid onto an
optical image from the HST. Copyright: MPIfR Bonn and \textit{Hubble
Heritage Team}. Graphics: \textit{Sterne und Weltraum} (from
\cite{fletcher+11}).} \label{fig:m51}
\end{figure}

\begin{figure}
\centering
\includegraphics[width=0.355\textwidth]{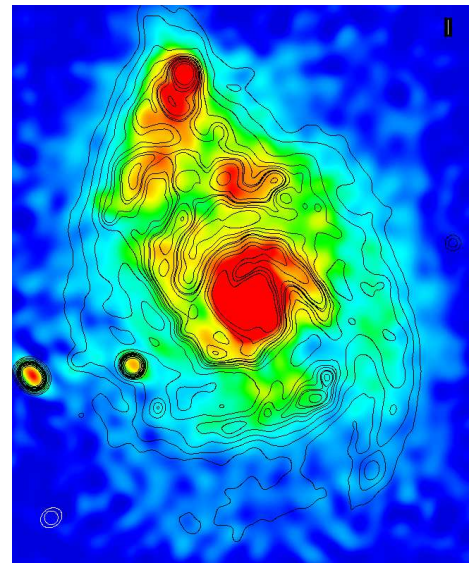}
\caption{Contours of total emission of M~51, observed at 1.4\,GHz
with the VLA (from \cite{fletcher11}), superimposed on a colour
image of the total radio intensity at 120--181\,MHz, observed with
LOFAR at 20\arcsec\ resolution (from David Mulcahy, MPIfR Bonn, and
the MKSP commissioning team).} \label{fig:m51_lofar}
\end{figure}

Turbulent as well as ordered magnetic fields have been detected in
the disks and halos of many galaxies (\cite{beck12a}). The field
lines mostly follow the material spiral arms in galaxies, e.g. in
M~51 (Fig.\ref{fig:m51}). The strongest regular fields are often
located between the arms, e.g. in NGC~6946 (\cite{beck07}).
Large-scale patterns of Faraday rotation measures (RM) are
signatures of coherent dynamo fields and can be identified from
polarized emission of the galaxy disks (\cite{fletcher11}) or from
RM data of polarized background sources (\cite{stepanov08}). Many
spiral galaxies host a dominating axisymmetric disk field, while
dominating bisymmetric fields are rare, as predicted by dynamo
models. Faraday rotation can be described in most galaxies by a
superposition of two or three azimuthal dynamo modes
(\cite{fletcher11}). In many galaxy disks no clear patterns of
Faraday rotation were found. Either the field structure cannot be
resolved with present-day telescopes or the generation of
large-scale modes takes longer than the galaxy's lifetime.

Important questions still remain regarding the amplification
process, as well as the configuration of the large-scale field in
evolving galaxies (\cite{moss12}) and the field structure in
galactic halos (\cite{braun10}). The expected number density of
background sources seen by LOFAR will enable systematic studies of
galactic field structures using Faraday rotation of background
sources (\cite{stepanov08}). ``Faraday spectra'' generated by RM
Synthesis (Sect.~\ref{rm}) will allow a detailed 3-D view of regular
magnetic fields and their reversals (\cite{bell11,frick11}) and
enable a clear measurement of magneto-ionic turbulent fluctuations
and their scale spectrum. These will give us a handle on the
properties of the turbulent motions responsible for dynamo action,
allowing us to address outstanding key questions: such as whether
magnetic fields are dynamically important in the ISM of galaxies at
different evolutionary stages. LOFAR's sensitivity to regions of low
gas density and weak field strengths will also allow us to measure
the magnetic structure in the outer disks and wider halos of spiral
galaxies. It is here that star formation activity is low, and
processes additional to dynamo action, such as gas outflows from the
inner disk, the magneto-rotational instability, gravitational
interaction and ram pressure by the intergalactic medium are
imprinted on the magnetic structure.

The first LOFAR maps of nearby galaxies (M~51 and NGC~4631) were
published by \cite{mulcahy12}. More recent observations show an
extended radio disk around M~51 (Fig.~\ref{fig:m51_lofar}).

\subsection{Magnetic fields in the radio galaxy M~87}
\label{m87}

\begin{figure}
\centering
\includegraphics[width=0.35\textwidth]{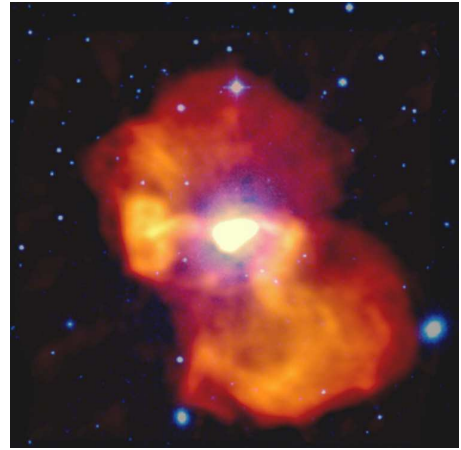}
\caption{Total synchrotron intensity of the radio galaxy M~87,
observed with LOFAR at 115--162\,MHz and overlaid onto an optical
SDSS image. The size of the shown field is 19\arcmin\ x 19\arcmin,
the resolution is 21\arcsec\ x 15\arcsec\ (from \cite{gasperin12}).}
\label{fig:m87}
\end{figure}

M~87 = Vir~A is one of the nearest radio galaxies, the fourth
brightest radio source in the northern sky and hosts in its nucleus
one of the most massive black holes discovered so far. The
low-frequency emission from M~87 forms a huge halo around the galaxy
(Fig.~\ref{fig:m87}). The similarity to the radio images at higher
frequencies indicates that the halo is confined by the pressure of
the intergalactic medium. The emitting cosmic-ray electrons in the
lobes are freshly provided by the active nucleus
(\cite{gasperin12}).

\subsection{Magnetic fields in galaxy clusters}
\label{clusters}

Some fraction of galaxy clusters, mostly the X-ray bright ones,
have, diffuse radio emission, emerging from diffuse halos and
steep-spectrum ``relic'' sources at the periphery of clusters. The
diffuse halo emission is unpolarized and emerges from intracluster
magnetic fields. In contrast to the interstellar medium, the
magnetic energy density is much smaller than that of the thermal gas
in the intracluster medium. Still, intracluster fields affect
thermal conduction in the halo gas, give rise to instabilities and
hence can modify the structure of the intracluster medium (see
Br\"uggen, this volume).

RMs towards background sources show a vanishing mean value and a
dispersion which decreases with distance from the cluster center
(\cite{clarke01,bonafede11}). The intracluster fields are turbulent
with correlation lengths of several 10\,kpc (\cite{murgia04}).

Relic sources can emit highly polarized radio waves from anisotropic
magnetic fields generated by compression in merger shocks
(\cite{ensslin98,iapichino12}). A polarized region of about 2\,Mpc
size was discovered in the cluster 1RXS J0603.3+4214 at a redshift
of 0.225 (\cite{weeren12b}).

\begin{figure}
\centering
\includegraphics[width=0.4\textwidth]{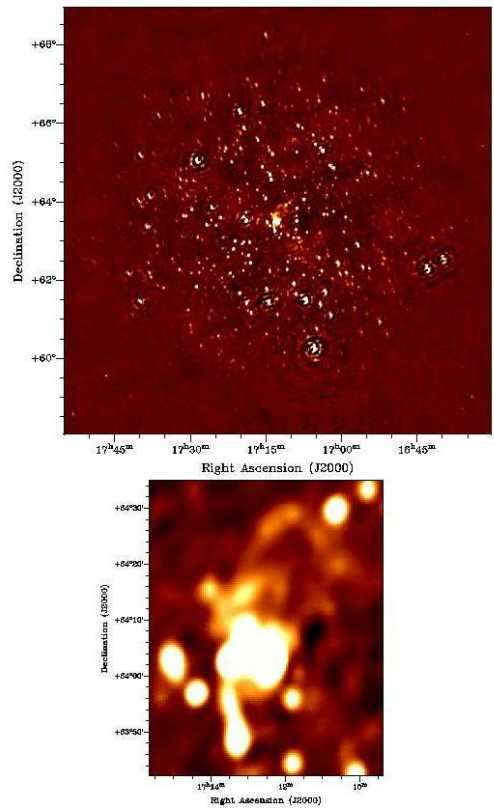}
\caption{Top panel: Total synchrotron intensity around the galaxy
cluster A~2255, observed with LOFAR at 110--190\,MHz. The field size
is 11\degr\ x 11\degr, the resolution is 2\arcmin. Bottom panel:
zoom into the inner 1\degr\ with the extended cluster halo, the
relic source at the right periphery of the halo and a radio galaxy
to the south (from Roberto Pizzo, ASTRON).} \label{fig:A2255}
\end{figure}

A~2255 is a nearby, rich cluster, which shows signs of undergoing a
merger event. It hosts a diffuse radio halo and a highly polarized
``relic'' source at its periphery (\cite{govoni05,pizzo09}). The
LOFAR image (Fig.~\ref{fig:A2255}) confirms the diffuse halo and the
relic source, and shows several radio galaxies in the neighborhood
of the cluster. Another giant cluster halo, around the nearby
cluster A~2256, was mapped with the LOFAR low band between 18\,MHz
and 67\,MHz (\cite{weeren12a}), to give the first image of a galaxy
cluster at frequencies below 30\,MHz.

\section{The Magnetism Key Science Project}
\label{mksp}

The Magnetism Key Science Project (MKSP) aims to exploit the unique
abilities of LOFAR to investigate cosmic magnetic fields in a
variety of astrophysical sources (\cite{anderson12}). At the end of
2012, the MKSP Project Team consisted of 86 members from 13
countries and is led by a German/Dutch/UK management team. MKSP
website:
\textit{http://www.mpifr-bonn.mpg.de/staff/rbeck/MKSP/mksp.html}

The MKSP Project Plan includes an initial target list of galaxies,
selected in close cooperation with the LOFAR Survey Key Science
Project (\cite{rottgering11}), to be followed by deep observations
of the diffuse total emission and the Faraday spectra from selected
regions in the Milky Way, a variety of nearby galaxies, selected
galaxy groups and the galaxies of the Virgo cluster. Sub-areas of
the deep fields centered on polarized background sources will also
be imaged with high resolution (about 1\arcsec) to obtain grids of
RM values derived from background sources. A minimum of about 10
background sources is needed to recognize a large-scale field
pattern in a galaxy (\cite{stepanov08}). High angular resolution is
needed to reduce depolarization within the telescope beam. The
international LOFAR stations provide baselines currently up to about
1300\,km, yielding sub-arcsecond resolution (\cite{heald11}). The
deep fields around galaxies will be located at different Galactic
latitudes and will also be analyzed with respect to the properties
of the small-scale magnetic field in the foreground of the Milky
Way, e.g. by computing the structure functions as a function of
Galactic latitude.

The structure of small-scale magnetic fields will be studied in the
lobes of giant radio galaxies. Polarized synchrotron emission and
rotation measures from pulsars and polarized jets from young stars
will be observed in cooperation with the Transients Key Science
Project (\cite{fender06,stappers11}). The radio polarization of
galaxy clusters will be studied by the LOFAR Survey Key Science
Project, in cooperation with the MKSP.

Detecting Faraday rotation signals from intergalactic magnetic
fields is a challenge which requires a very large number density of
sources and hence a very high sensitivity. Studies of diffuse
emission from intergalactic filaments will use the LOFAR surveys
that include the Coma field, followed by deeper observations.
Primary targets are a $\approx$\,100 sq.~degree area, centered on
the Coma cluster of galaxies, and compact galaxy groups with minimum
Galactic foreground contribution. Proof for an intergalactic origin
of specific components of the Faraday spectrum could come from a
statistical comparison with source redshifts.

\section{Outlook: the Square Kilometre Array}
\label{ska}

High-resolution, deep observations at high frequencies, where
Faraday depolarization is small, require a major increase in
sensitivity for continuum observations, which will be achieved by
the planned Square Kilometre Array (SKA, {\em www.skatelescope.org})
(\cite{garrett10,beck12b}). The detailed structure of the magnetic
fields in the ISM of galaxies, in galaxy halos, in cluster halos and
in cluster relics can then be observed. Direct insight into the
interaction between gas and magnetic fields in these objects will
become possible. The SKA will also allow us to measure the Zeeman
effect in much weaker magnetic fields in the Milky Way and in nearby
galaxies.

Detection of polarized emission from distant, unresolved galaxies
will reveal large-scale ordered fields, and statistics can be
compared with the predictions of dynamo theory (\cite{arshakian09}).
The SKA at 1.4~GHz will detect Milky-Way type galaxies at about
$z\le1.5$ (\cite{murphy09}) and their polarized emission at about
$z\le0.5$. Bright starburst galaxies can be observed at larger
redshifts, but are not expected to host ordered fields. Cluster
relics are also detectable at large redshifts through their
integrated polarized emission. Unpolarized synchrotron emission,
signature of turbulent magnetic fields, can be detected with the SKA
out to very large redshifts in starburst galaxies, depending on
luminosity and magnetic field strength, and also in cluster halos.

If polarized emission from galaxies, cluster halos or cluster relics
is too weak to be detected, the method of RM grids towards
background QSOs can still be applied and will allow us to determine
the field strength and pattern in an intervening object. Here, the
distance limit is given by the polarized flux of the background QSO
which can be much higher than that of the intervening galaxy.
Regular fields of several $\mu$G strength were already detected in
distant galaxies (\cite{bernet08,kronberg08}). Mean-field dynamo
theory predicts RMs from evolving regular fields with increasing
coherence scale at $z\le3$ (\cite{arshakian09}). A reliable model
for the field structure of nearby galaxies, cluster halos and
cluster relics needs RM values from a large number of polarized
background sources, hence large sensitivity and/or high survey
speed. POSSUM, the RM survey around 1\,GHz with the planned
Australia SKA Pathfinder (ASKAP) telescope with 30\,deg$^2$ field of
view, will measure about 100 RM values from polarized extragalactic
sources per square degree.

The SKA ``Magnetism'' Key Science Project plans to observe a
wide-field survey (at least $10^4$~deg$^2$) around 1\,GHz with 1\,h
integration per field which will measure at least 1500 RMs
deg$^{-2}$, or at least $2\times10^7$ RMs at a mean spacing of
$\simeq90\arcsec$ (\cite{gaensler04}). More than 10\,000 RM values
are expected in the area of M~31 and will allow the detailed
reconstruction of the 3-D field structure in this and many other
nearby galaxies, while simple patterns of regular fields can be
recognized out to distances of about 100\,Mpc (\cite{stepanov08}).
The magnetism of cluster halos can be measured by the RM grid to
redshifts of about 1 (\cite{krause09}). Finally, the SKA pulsar
survey should find more than 10\,000 new pulsars, which will be
mostly polarized and reveal RMs, suited to map the Milky Way's
magnetic field with high precision (\cite{noutsos12}).

Construction of the SKA is planned to start in 2017. In the first
phase (until about 2020) about 10\% of the SKA will be erected
(SKA$_1$), with completion of construction at the low and mid
frequency bands by about 2025 (SKA$_2$), followed by construction at
the high band.

\acknowledgements The LOFAR commissioning teams are acknowledged for
performing a great job. James Anderson is thanked for careful
reading the manuscript. RB, AH and DM acknowledge financial support
from DFG Research Unit 1254.\\
LOFAR, designed and constructed by ASTRON, has facilities in several
countries, that are owned by various parties (each with their own
funding sources), and that are collectively operated by the
International LOFAR Telescope (ILT) foundation under a joint
scientific policy.

\end{document}